\documentstyle[preprint,prb,aps,eqsecnum,epsf,tighten]{revtex}

\begin{document}
\draft
\title{Berry phase, hyperorbits, and the Hofstadter spectrum \\
---  semiclassical dynamics in magnetic Bloch bands}
\author{Ming-Che~Chang and Qian~Niu}
\address{Department of Physics, University of Texas at Austin, Austin, TX
78712}
\date{\today}
\maketitle
\begin{abstract}
{We have derived a new set of semiclassical equations for electrons in magnetic
Bloch bands.
The velocity and energy of magnetic Bloch electrons are found to be modified
by the Berry phase and magnetization.
This semiclassical approach is used to study general electron transport
in a DC or AC electric field. We also find a close connection between the
cyclotron orbits in magnetic Bloch bands and the energy subbands in the
Hofstadter spectrum.
Based on this formalism, the pattern of band splitting, the distribution of
Hall
conductivities, and the positions of energy subbands in the Hofstadter spectrum
can
be understood in a simple and unified picture.}
\end{abstract}
\narrowtext

\section{Introduction}

The semiclassical method has played a very important role in
studying electron dynamics in periodic systems.\cite{Mermin} In this approach,
the effect of
a periodic potential is treated by quantum mechanical methods and yields usual
band structure for energy spectrum, while an extra electromagnetic field is
treated as
a classical perturbation.
The velocity of an electron in the one-band approximation is given by
\begin{equation}
{\bf\dot r}=\frac{\partial{\cal E}_n({\bf k})}{\hbar\partial {\bf k}},
\end{equation}
where ${\cal E}_n$ is the energy spectrum for the $n$-th band.
The dynamics of quasi-momentum ${\bf k}$ is governed by the Lorentz-force
formula
\begin{equation}
\hbar{\bf \dot k}=-e{\bf E}-e{\bf \dot r}\times {\bf B},
\end{equation}
where ${\bf E}$ and ${\bf B}$ are the external electric and magnetic fields.
These equations may be regarded  as the equations of motion for the center of
mass of a wave
packet in the ${\bf r}$ and ${\bf k}$-spaces. Tremendous amount of work has
been done to
justify these simple looking formulae and to their quantization.\cite{history}

These formulae, however,  become invalid if the magnetic field is so strong
that it is no longer appropriate to be treated as a perturbation.
In this case, we need to solve the Schrodinger equation with the following
Hamiltonian:
\begin{equation}
H_0=\frac{1}{2m}\left( -i\hbar\frac{\partial}{\partial {\bf r}}+
e{\bf A}_0({\bf r})\right)^2+V({\bf r}),
\end{equation}
where ${\bf A}_0({\bf r})$ is the vector potential of a homogeneous
magnetic field,\cite{Changmm} and $V({\bf r})$ is a periodic potential.
The eigen-energies of Eq.~(1.3) will be called magnetic Bloch bands, and its
energy
eigenstates, magnetic Bloch states. A crucial difference between
a Bloch state and a magnetic Bloch state lies in their translational
properties.
The Hamiltonian $H_0$ is not invariant under lattice translation because ${\bf
A}_0({\bf r})$
cannot be a periodic function if the mean value of ${\bf B}$ is not zero.
However, $H_0$ can be made
invariant under ``magnetic" translation operators, which are the usual
translation operators
multiplied by a position dependent phase factor.\cite{Zak}

We first give  a brief review of the magnetic translation symmetry.
In order to simplify the discussion, we assume the motion of electrons is
confined in a plane (${\bf r}=(x,y)$) and the magnetic field is along the $z$
direction.
A magnetic Bloch state is the state that satisfies
\begin{equation}
H_0\Psi_{n{\bf k}}({\bf r})={\cal E}_n({{\bf k}})\Psi_{n{\bf k}}({\bf r}),
\end{equation}
as well as
\begin{eqnarray}
{\tilde T}_1(R_1)\Psi_{n{\bf k}}({\bf r})&=&e^{ik_1R_1}\Psi_{n{\bf k}}({\bf
r})\cr
{\tilde T}_2(R_2)\Psi_{n{\bf k}}({\bf r})&=&e^{ik_2R_2}\Psi_{n{\bf k}}({\bf
r}),
\end{eqnarray}
where ${\tilde T_1}$ and ${\tilde T_2}$ are magnetic translation operators.
Although ${\tilde T_1}$ and ${\tilde T_2}$ commute with the Hamiltonian  by
construction,
they do not commute with each other
unless there is an integer number of flux
quantum $\phi_0$ enclosed by $|{\bf R}_1\times{\bf R}_2|$.
Therefore, when the magnetic flux is a rational multiple $p/q$ of the flux
quantum $\phi_0$
per unit cell of the lattice (plaquette), we must choose a ``magnetic" unit
cell containing $q$ plaquettes in order that both $k_1$ and $k_2$ be
good quantum numbers. $\Psi_{n{\bf k}}$ thus defined forms a complete set and
satisfies the orthogonality condition
\begin{equation}
\langle \Psi_{n'{\bf k}'}|\Psi_{n{\bf k}}\rangle=\delta_{n'n}\delta_{{\bf
k}'{\bf k}}.
\end{equation}
The domain of ${\bf k}$ is a magnetic Brillouin zone (MBZ), which is
$q$ times smaller than a usual Brillouin zone. Furthermore, because of the
magnetic
translation symmetry, the MBZ has exactly a $q$-fold degeneracy.
We will call each repetition unit a ``reduced" MBZ.

One example of magnetic Bloch bands is the subbands split from Bloch bands
due to a magnetic field.
The number of subbands into which a band splits
depends on the magnetic flux per plaquette in an
intricate way.\cite{Harper} If $\phi=p/q$ (in units of $\phi_0$),
a Bloch band will split into $q$ magnetic subbands. On the other hand, if the
magnetic
field is very strong, it is more appropriate to treat the lattice
potential as a perturbation, then a Landau level
will be broadened and split into $p$ subbands.

For a usual solid with a lattice constant $a=5{\rm \AA}$, the magnetic field
has
to be as large as $10^4$ Tesla in order for $p/q$ to be of order unity. This
is the reason the splitting was once considered impossible to observe.
However, the field strength can be greatly reduced to a few
Tesla if we use an artificial lattice with a much larger
(say, $500{\rm \AA}$) lattice constant.
Evidence for such splitting
has appeared in recent transport measurements.\cite{evidence}
It is expected that more evidence will emerge in the future
by using a very pure sample in a very low temperature environment.
Under such circumstances, what is the dynamics for electrons in such magnetic
Bloch bands?

Using the magnetic Bloch states as an unperturbed basis,
we found the following semiclassical dynamics in magnetic Bloch
bands:\cite{Changprl}
\begin{equation}
{\bf{\dot r}}=\frac{\partial E_n({\bf k})}{\hbar\partial
{\bf k}}-{\bf\dot k}\times {\bf \Omega}_n({\bf k}),
\label{main1}
\end{equation}
and
\begin{equation}
\hbar{\bf\dot{k}}=-e{\bf E}-e{\bf\dot{r}}\times\delta {\bf B},
\label{main2}
\end{equation}
where $E_n({\bf k})$
consists of a band energy ${\cal E}_n({\bf k})$ and a correction from the
magnetic
moment of the wave packet (this correction did not appear in Ref.~7).
${\bf \Omega}_n({\bf k})$ is the ``Berry curvature", whose integral over
an area bounded by a path $C$
in ${\bf k}$-space is the Berry phase $\Gamma_n(C)$.\cite{Berry}
${\bf E}$ and $\delta {\bf B}$ are
external fields added to the already present $B_0$ field.
These equations will be derived and explained in detail in Sec.~II.

Despite the similarities between Eqs.~(1.1), (1.2) and Eqs.~(\ref{main1}),
(\ref{main2}), there are
several essential differences.
See Table~I for a comparison between this new semiclassical dynamics and the
conventional one.
The last item in Table~I, about the quantization of orbits, will be explained
in Sec.~IV.
We have to emphasize that the $\delta B$ in Eq.~(\ref{main2})
is the field applied to the magnetic Bloch states;
it is not the total magnetic field applied to the sample.
This separation is particularly useful when $B(x,y)$ is composed of a large
constant part $B_0$ and a small non-uniform part $\delta B(x,y)$. In this case,
we
can calculate the effect of $B_0$ exactly and treat $\delta B(x,y)$
as a classical perturbation.

This paper is organized as follows:
Sec.~II is devoted to the derivation of Eqs.~(\ref{main1}) and (\ref{main2}).
Their use in calculating transport properties in a DC or AC electric field
is demonstrated in Sec.~III.  The presence of $\delta B$ will lead to formation
of cyclotron
orbits in magnetic Bloch bands, similar to the formation of cyclotron orbits
in usual Bloch bands. This is explained in Sec.~IV. In Sec.~V, we explore
the connection between these cyclotron orbits
and the Hofstadter spectrum.
In Sec.~VI, we estimate the energy levels in the Hofstadter
spectrum by calculating the cyclotron energies in magnetic Bloch bands.
Finally, this paper is summarized in Sec.~VII.

\section{Derivation  of the new semiclassical dynamics}

The method we use is to construct a wave packet out of $\Psi_{n{\bf k}}$ (hence
it is already
included the effect of $B_0$), and study its motion governed by the
following Hamiltonian:
\begin{equation}
H=\frac{1}{2m}\left[ -i\hbar\frac{\partial}{\partial {\bf r}}+
e{\bf A}_0({\bf r})+e\delta{\bf A}({\bf r},t)\right]^2+V({\bf r}),
\end{equation}
where $-\partial \delta{\bf A}/\partial t={\bf E}$,
and $\nabla\times\delta {\bf A}=\delta {\bf B}$.
For simplicity we assume both ${\bf E}$ and $\delta{\bf B}$ are uniform; the
derivation is still valid if they are slowly varying in space and/or time.

\subsection{Wave packet in a magnetic Bloch band}

Our derivation will be confined to one energy band by neglecting interband
transitions;
therefore, the band index $n$ is henceforth dropped.
Consider the following wave packet centered at ${\bf r}_c$ which is
formed from the superposition of magnetic Bloch states,
\begin{equation}
|W_0\rangle=\int_{\rm MBZ} d^2{\bf k} ~w({\bf k})~|\Psi({\bf k})\rangle,
\end{equation}
where $w({\bf k})$ is a function localized around ${\bf k}_c$ (see Fig.~1 for
an illustration).
It has to be chosen such that
\begin{equation}
\int d^2{\bf k} ~{\bf k} ~|w({\bf k})|^2={\bf k}_c,
\label{centek}
\end{equation}
and
\begin{equation}
\langle W_0|{\bf r}|W_0\rangle ={\bf r}_c.
\label{center}
\end{equation}
By defining $u_{{\bf k}}({\bf r})= e^{-i{\bf k}\cdot{\bf r}}\Psi_{{\bf k}}({\bf
r})$,
the mean position of $W_0$ can be written as
\widetext
\begin{eqnarray}
\langle W_0|{\bf r}| W_0\rangle &=& \int d^2{\bf k}' \int d^2 {\bf k}\
w^*({\bf k}')w({\bf k}) \langle \Psi({\bf k}')|
\left(-i\frac{\partial}{\partial {\bf k}}e^{i{\bf k}\cdot{\bf r}}
\right)|u({\bf k})\rangle \cr
&=& \int d^2{\bf k}' \int d^2 {\bf k} \ w^*({\bf k}')w({\bf k})
\left[ \left(-i\frac{\partial}{\partial {\bf k}}
\right) \delta({\bf k}-{\bf k}') + \delta({\bf k}-{\bf k}') \langle u({\bf
k})|i\frac{\partial}{\partial {\bf k}}
|u({\bf k})\rangle_{\rm cell} \right] \cr
&=&\int d^2{\bf k}\left[ w^*({\bf k})i\frac{\partial}{\partial {\bf k}}w({\bf
k})+|w({\bf k})|^2
\langle u({\bf k})|i\frac{\partial}{\partial {\bf k}}|u({\bf k})\rangle_{\rm
cell}\right],
\end{eqnarray}
\narrowtext
\begin{figure}
\epsfxsize=3in
\hskip 1.0truein \epsffile{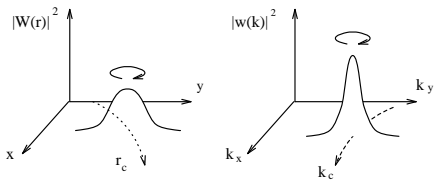}
\caption{Schematic plots for the motion of wave packets in ${\bf
r}$ and ${\bf k}$- space.}
\end{figure}
where we have used the identity\cite{Blount}
\begin{equation}
\langle u({\bf k}')|e^{i({\bf k}-{\bf k}')\cdot {\bf
r}}~i\frac{\partial}{\partial {\bf k}}|u({\bf k})\rangle
=\delta({\bf k}-{\bf k}')\langle u({\bf k})|i\frac{\partial}{\partial {\bf
k}}|u({\bf k})\rangle_{\rm cell}.
\end{equation}
The subscript means that the spatial integration is restricted to a magnetic
unit cell.
By defining
\begin{equation}
\mbox{\boldmath${\cal A}$}({\bf k})= i\langle u({\bf k})
|\frac{\partial}{\partial {\bf k}}| u({\bf k})
\rangle_{\rm cell},
\end{equation}
Eq.~(\ref{center}) can be written as
\begin{equation}
\int d^2{\bf k} \left[ w^*({\bf k})i\frac{\partial}{\partial {\bf k}}w({\bf
k})+|w({\bf k})|^2
\mbox{\boldmath${\cal A}$}({\bf k}) \right] ={\bf r}_c.
\label{condition}
\end{equation}

\subsection{Effective Lagrangian for a moving wave packet}

The dynamics of
a moving wave packet is governed by the following effective Lagrangian
\begin{equation}
L({\bf r}_c,{\bf k}_c,{\bf{\dot r}_c},{\bf{\dot k}_c})=\langle W|
i\hbar\frac{\partial}{\partial t}| W\rangle-\langle W| H |W\rangle,
\end{equation}
where $W$ is a wave packet centered at ${\bf r}_c$ and ${\bf k}_c$
in the presence of external electromagnetic fields (${\bf k}_c$ is treated as a
generalized coordinate here).
We can always choose a gauge such that the vector potential
$\delta {\bf A}$ is locally gauged away at a chosen point ${\bf r}={\bf r}_c$.
At this particular point, the moving wave packet $W$
is the same as the $W_0$ in Eq.~(2.2).\cite{gauge}
The value of $W$ near ${\bf r}_c$ can be approximated as
\begin{equation}
W({\bf r})=e^{-ie/\hbar \delta {\bf A}({\bf r}_c,t)\cdot {\bf r}}W_0({\bf r}).
\end{equation}
First, we evaluate the energy of this wave packet, which is
$\langle W| H | W\rangle= \langle W_0| H' |
W_0\rangle$, with
\begin{eqnarray}
H' &=&\frac{1}{2m}\left\{ -i\hbar\frac{\partial}{\partial {\bf r}}+
e{\bf A}_0({\bf r})+e\left[\delta{\bf A}({\bf r},t)-\delta{\bf A}({\bf
r}_c,t)\right]
\right\}^2+V({\bf r}) \cr
&\simeq& H_0+\frac{e}{2m}\left\{
\left[\delta{\bf A}({\bf r},t)-\delta{\bf A}({\bf r}_c,t)\right]\cdot {\bf
P}+{\rm h.c.}\right\}.
\end{eqnarray}
${\bf P}$ is the mechanical momentum operator corresponding to $H_0$.
For simplicity, we choose the circular gauge for $\delta B$, which gives
$\delta {\bf A}({\bf r},t)=
-{\bf E}t+\frac{1}{2}\delta {\bf B}\times {\bf r}$, and leads to
\begin{equation}
H' \simeq H_0+\frac{e}{2m}\delta {\bf B}\cdot {\bf L},
\end{equation}
where ${\bf L}=({\bf r}-{\bf r}_c)\times {\bf P}$ is the mechanical angular
momentum
of the wave packet about its center of mass.
Therefore,
\begin{equation}
\langle W| H |W \rangle \simeq {\cal E}({\bf k}_c) +\frac{e}{2m}\delta {\bf
B}\cdot
\langle W_0|{\bf L}| W_0\rangle.
\label{eq1}
\end{equation}
The second term represents the energy correction due to magnetic
moment of the wave packet.
Notice that while a wave packet in an ordinary Bloch band does not rotate,
a wave packet in a magnetic Bloch band usually does.

For the first term on the right hand side of Eq.~(2.9), we have
\begin{eqnarray}
&\ &\langle W_0|i\hbar\frac{\partial}{\partial t} |W_0\rangle\cr
&=&\langle W_0|e\delta{\bf \dot A}({\bf r}_c,t)\cdot{\bf r}|W_0\rangle
+\int d^2{\bf k}~w^*({\bf k}) i\hbar\frac{\partial}{\partial t} w({\bf k}) \cr
&=& e\delta{\bf\dot A}({\bf r}_c,t)\cdot {\bf r}_c
+\int d^2 {\bf k}~|w({\bf k})|^2\hbar\frac{\partial}
{\partial t} \gamma({\bf k},t) \cr
&\simeq& e\delta {\bf \dot A}({\bf r}_c,t)\cdot {\bf r}_c
+\hbar\frac{\partial}{\partial t}\gamma({\bf k}_c,t),
\end{eqnarray}
where $w({\bf k})\equiv |w({\bf k})|e^{-i\gamma({\bf k},t)}$. Up to terms of
total time derivative, which
have no effect on the dynamics, the last line can be written as
\begin{equation}
-e\delta {\bf A}\cdot{\bf \dot r}_c-\hbar {\bf \dot k}_c\cdot \frac{\partial}
{\partial {\bf k}}_c\gamma({\bf k}_c,t).
\label{first}
\end{equation}
Using the condition in Eq.~(\ref{condition}) and neglecting another term of
total time
derivative, we can write Eq.~(\ref{first}) as
\begin{equation}
-e\delta {\bf A}\cdot{\bf \dot r}_c
+\hbar{\bf k}_c\cdot {\bf \dot r}_c+\hbar{\bf \dot k}_c
\cdot {\mbox{\boldmath${\cal A}$}}({\bf k}_c).\label{eq2}
\end{equation}

Combining Eqs.~(\ref{eq1}) and (\ref{eq2}), we have
a final form for the effective Lagrangian (omitting subscript $c$)
\begin{equation}
L({\bf r},{\bf k},{\bf \dot r},{\bf \dot k})=
-e\delta{\bf A}({\bf r},t)\cdot{\bf \dot r}+\hbar{\bf k}\cdot{\bf\dot r}+
\hbar \mbox{\boldmath${\cal A}$}({\bf k})\cdot {\bf\dot {\bf k}}-E({\bf k}),
\end{equation}
where $E({\bf k})\equiv {\cal E}({\bf k})+(e/2m)\delta {\bf B}\cdot {\bf
L}({\bf k})$.
Under a gauge transformation for ${\bf A}_0({\bf r})$,
the Berry potential $\mbox{\boldmath${\cal A}$}({\bf k})$ will be changed by a
term like
$\partial_{{\bf k}_0} \chi({\bf k}_0)$, and this will only
change $L$ by a total time derivative. This is also true if a different
gauge is chosen for $\delta {\bf B}$. Therefore, the dynamics
is invariant under gauge transformation.

The dynamical equations in Eqs.~(\ref{main1}) and (\ref{main2}) can be obtained
straightforwardly from this Lagrangian by using the Euler-Lagrange
equation. The relation between ${\bf \Omega}({\bf k})$
and $\mbox{\boldmath${\cal A}$}({\bf k})$ is
\begin{equation}
{\bf \Omega}({\bf k})=\nabla\times \mbox{\boldmath${\cal A}$}({\bf k}),
\end{equation}
which can also be written as ($z$-component)
\begin{equation}
\Omega({\bf k})=i\left( {\langle \frac{\partial u_n}{\partial k_1}|
\frac{\partial u_n}{\partial k_2}\rangle-
\langle \frac{\partial u_n}{\partial k_2}
|\frac{\partial u_n}{\partial k_1}\rangle } \right).
\label{master}
\end{equation}
This is the familiar Berry curvature in the study of the quantum Hall
effect.\cite{TKNdN}

\section{Transport in magnetic Bloch bands}

\subsection{Transport by an electric field}

The next step is to combine the semiclassical equations with Boltzmann
equation to study the transport properties of magnetic Bloch electrons. The
Boltzmann equation is
\begin{equation}
{\bf{\dot r}}\cdot \frac{\partial f}{\partial {\bf r}}+
{\bf{\dot k}}\cdot \frac{\partial f}{\partial {\bf k}}=
\left( \frac{\partial f}{\partial t}\right)_{\rm coll},
\end{equation}
where $f=f({\bf r},{\bf k})$ is a distribution function.
The effect of impurities is included in the collision term
$\left( \partial f/\partial t\right)_{\rm coll}$.
We use the relaxation time approximation to replace it
by $-(f-f_0)/\tau({\bf k})$.
For a random distribution of delta impurities $v_0\sum_i \delta({\bf r}-{\bf
r}_i)$,
the impurity scattering rate is
\begin{equation}
\tau({\cal E}_F)^{-1}=\frac{\pi}{\hbar}\rho({\cal E}_F)n_iv_0^2,
\end{equation}
where $n_i$ is the area density of impurities.
This rate is proportional to the density of states $\rho({\cal E}_F)$ at Fermi
energy,
which varies wildly
with ${\cal E}_F$ because the energy spectrum is discrete.
However, since the following calculation is confined to only one band,
$\tau$ will be approximated by a constant.

The equations of motion of an electron subject to a uniform electric field are
\begin{equation}
{\bf{\dot r}}=\frac{\partial {\cal E}({\bf k})}{\hbar\partial
{\bf k}}+\frac{e}{\hbar}{\bf E}\times {\bf \Omega}({\bf k}),\ \ {\bf\dot
k}=-e{\bf E},
\end{equation}
where ${\cal E}({\bf k})$ is the reduced form of $E({\bf k})$ in the absence of
$\delta B$.
Substituting the expressions for ${\bf \dot r}$ and ${\bf \dot k}$ in Eq.~(3.3)
into
Eq.~(3.1), and setting $f=f_0$ on the left hand side of Eq.~(3.1), we obtain
\begin{equation}
f=f_0-\tau\left( \frac{\partial {\cal E}}{\hbar\partial {\bf k}}\cdot
\frac{\partial f_0}{\partial {\bf r}}+\frac{e}{\hbar} ({\bf E}\times {\bf
\Omega})
\cdot \frac{\partial f_0}{\partial {\bf r}}-
\frac{e}{\hbar}{\bf E}\cdot \frac{\partial f_0}{\partial {\bf k}} \right),
\end{equation}
where $f_0$ is the distribution function in equilibrium.
Electric current is given by
\begin{equation}
{\bf J}=-e\int \frac{d^2 {\bf k}}{(2\pi)^2} f{\bf\dot r}.
\end{equation}
It can be decomposed into three parts, ${\bf J}^\Omega+{\bf J}^\tau+{\bf
J}^{\mu}$,
with the following definitions:
\begin{eqnarray}
{\bf J}^\Omega &=& -{\bf E} \times
\frac{e^2}{\hbar}\int \frac{d^2{\bf k}}{(2\pi)^2}f_0{\bf \Omega}({\bf k}),\cr
{\bf J}^\tau &=& e^2 \tau \int \frac{d^2{\bf k}}{(2\pi)^2}
\left(-\frac{\partial f_0}{\partial {\cal E}}\right){\bf v_b}\left({\bf
v_b}\cdot
{\bf E}\right ),\cr
{\bf J}^{\mu} &=& e\tau \int \frac{d^2 {\bf k}}{(2\pi)^2}
\left(-\frac{\partial f_0}{\partial {\cal E}}\right){\bf v}\left({\bf v}\cdot
\frac{\partial \mu}{\partial {\bf r}}\right).
\end{eqnarray}
In these expressions, ${\bf v_b}$ is the velocity $\partial {\cal
E}/\hbar\partial {\bf k}$ due to
energy dispersion, and
${\bf v}$ is the total velocity in Eq.~(3.3). The meaning of these currents
is explained below.

First, ${\bf J}^{\Omega}$ is the Hall current. This is most evident considering
a filled band with $f_0=1$. In this case,
both ${\bf J}^\tau$ and ${\bf J}^\mu$
vanish, and only ${\bf J}^\Omega$ is nonzero.
The integral of $\Omega({\bf k})$ over one magnetic Brillouin zone divided by
$2\pi$ is always an integer,
which is the topological Chern number discovered by Thouless {\it et al.}
\cite{TKNdN} Therefore we have
\begin{equation}
{\bf J}^\Omega=-{\rm C}\frac{e^2}{h}{\bf E}\times {\hat z},\ \ ({\rm C}\in Z).
\end{equation}
This formula represents the quantization of Hall current for a magnetic
Bloch band.

Second, ${\bf J}^\tau$ is the diffusion current due to disorder scatterings.
It can be put in the following form
\begin{equation}
{\bf J}^\tau=e^2\int \frac{d^2{\bf k}}{(2\pi)^2}\left(-\frac{\partial
f_0}{\partial
{\cal E}} \right)\int^{\infty}_0 dt~{\bf v_b}(t){\bf v_b}(0)
\cdot {\bf E},
\end{equation}
where ${\bf v_b}(t)\equiv e^{-t/\tau}{\bf v_b}(0)$ is the current relaxed by
scatterings
after time $t$.  At very low temperature, it can be simplified to
\begin{equation}
{\bf J}^\tau=e^2g({\cal E}_F) {\bf\sf D}\cdot {\bf E},
\end{equation}
where $g({\cal E}_F)$ is the density of states at Fermi energy.
${\bf\sf D} \equiv \int_0^\infty dt \langle {\bf v_b}(t) {\bf v_b}(0) \rangle$
is the
diffusion tensor, and the angular
bracket $\langle \ \rangle$ means averaging over the Fermi surface.
\cite{Beenakker}

Third, ${\bf J}^\mu$ is the current due to density gradient. It can be
put in a form similar to ${\bf J}^{\tau}$ in Eq.~(3.9) at low temperature, but
with two changes:
(1) The velocity ${\bf v_b}$ in ${\bf\sf D}$ is replaced by the total velocity
${\bf v}$ that includes the curvature term. (2) The driving force
$e{\bf E}$ is replaced by $\partial \mu/\partial {\bf r}$.
To evaluate this current, we need to know the explicit form of $\Omega({\bf
k})$.
This in general requires numerical calculation (see Sec.~VI).

We remark that, even though the derivation of Eq.~(3.8) is based on a uniform
electric field,
its validity goes beyond that. It is actually a kinetic formulation,
first proposed by Chambers,
that is also valid in the presence of a magnetic field
(for magnetic Bloch bands, the magnetic field is $\delta B$).\cite{Chambers}

\subsection{Perturbation by an AC electric field}

The semiclassical method is much simpler to use than
full quantum mechanical approaches. This is most evident when the
perturbation is changing in time. We illustrate this by considering a
magnetic Bloch electron in an AC electric field.
To simplify the discussion, we will neglect the effect of disorder
and focus on the dynamics itself.

Assuming that a uniform electric field along the $x$ direction oscillates
with a low frequency $\omega$, we then have
\begin{eqnarray}
{\bf{\dot r}}&=&\frac{\partial {\cal E}({\bf k})}{\hbar\partial
{\bf k}}-\frac{e}{\hbar}E_0e^{i\omega t}\Omega({\bf k}){\hat y},\cr
\hbar{\bf\dot{k}}&=&-eE_0e^{i\omega t}{\hat x}.
\end{eqnarray}
Considering a square lattice with the following energy spectrum
\begin{equation}
{\cal E}({\bf k})=2\left[\cos (k_1 a)+\cos (k_2 a) \right],
\end{equation}
and substituting the solution ${\bf k}(t)$ into the ${\bf\dot r}$ equation in
(3.10),
we have
\begin{eqnarray}
\hbar {\dot x}&=& 2a\sin \left(\frac{eE_0a}{\hbar\omega}\sin(\omega t)
-k_{0x}a\right) \cr
\hbar {\dot y}&=&-2a\sin(k_{0y} a)-\frac{eE_0}{\hbar}
\Omega({\bf k}(t))\cos (\omega t),
\end{eqnarray}
where $(k_{0x},k_{0y})$ is the initial value of ${\bf k}$.
It is not difficult to see that after many cycles of oscillation, there
is a net drift along the $x$ direction with average velocity
\begin{equation}
\langle {\dot x}\rangle=-\frac{2a}{\hbar}\sin(k_{0x} a)
J_0(z),
\end{equation}
where $J_0$ is the zeroth order Bessel function, and
$z\equiv eE_0a/\hbar\omega$ is a ratio between two energy scales.\cite{odds}
It can be seen that the original band transport velocity
$-2a/\hbar\sin(k_{0x}a)$
is modified by
$J_0(z)$ because of the AC field. The electron is immobile along the $x$
direction when $z$ is a zero of the Bessel function. This resembles
the collapse of the usual Bloch band in the AC Wannier-Stark ladder
problem.\cite{Holthaus}

Finally, we comment that the use of semiclassical equations is based on the
assumption that impurities do not alter the band
structure. Therefore, the electron dynamics between collisions can be nicely
described by
Eqs.~(\ref{main1}) and (\ref{main2}).
This assumption is no longer valid when $q$ is large. In that case,
disorder broadening tends to merge the subbands and wash out the fine
structure.
(This will be clearer after the discussion of hierarchical structure of the
energy
bands in Sec.~V.)
However, it was found that despite the energy spectrum has singular $B$
dependence,
the density of states appears as a continuous function of the magnetic field.
\cite{Gerhardts} Therefore, we divide total magnetic field $B$ into $B_0$ and
$\delta B$, where $B_0$ is related to the band structure undestroyed by
disorder,
and $\delta B$ is a small perturbation.  In this case, the semiclassical
dynamics in the
magnetic Bloch band of $B_0$, driven by $E$ and $\delta B$, will be employed in
the Boltzmann equation.

\section{Magnetic perturbation and hyperorbits}

The usual Bloch electron will circulate around
the Fermi surface along a constant energy contour
in the presence of a magnetic field. It is well-known that
the quantization of cyclotron orbits leads to the famous de Haas-Van Alphen
effect.
In this section, we study a similar type of
cyclotron motion in magnetic Bloch bands. It will be seen that
this investigation yields very fruitful results. In particular, it offers a
very simple
explanation for the complex Hofstadter spectrum, which will be shown in Sec.~V
and
Sec.~VI.

\subsection{General properties of hyperorbits}

Combining the two equations in (3.3), we can eliminate ${\bf \dot r}$ to obtain
\begin{equation}
\hbar{\bf\dot k}=-eZ_{\delta B}({\bf k})
\frac{\partial E({\bf k})}{\hbar\partial{{\bf k}}}
\times\delta {\bf B}.
\end{equation}
where $Z_{\delta B}({\bf k})\equiv (1+\Omega({\bf k})\delta Be/\hbar)^{-1}$ is
a
curvature correction factor. This equation determines the trajectories of
magnetic Bloch electrons in ${\bf k}$-space.
It is not difficult to see that ${\bf k}$ moves
along a constant energy contour of $E({\bf k})$ (which is slightly different
from ${\cal E}({\bf k})$).
In a classical picture, it is the drifting-center trajectory of
the tighter cyclotron orbit formed from $B_0$.
However, we have to emphasize that, the existence
of hyperorbits is of quantum origin and cannot be explained classically.
To differentiate them from the usual orbits of Bloch electrons, we will call
them 'hyperorbits'.
\cite{Pippard}
The hyperorbit in real space is derived from ${\bf \dot r}={\bf \dot k}\times
{\hat z}(\hbar/
e\delta B)$, which is the ${\bf k}$-orbit rotated by $\pi/2$
and scaled by the factor $\hbar/e\delta B$. It is also possible to define an
effective
cyclotron mass according to its frequency. However, this frequency will be very
sensitive to its energy if the magnetic Bloch band is narrow, which is usually
the case.

There are several ways to verify the existence of hyperorbits. One way
is through the measurement of magnetoresistance oscillation that originates
from the
quantization of hyperorbits.
This oscillation has a much shorter period than the usual
de Haas-van Alphen oscillation because the effective  magnetic field $\delta B$
is
much smaller.
The other way of verifying it is by using an electron focusing device to detect
its
real space orbit.\cite{van Houten}
This method has been used  to map out the shape of a Fermi
surface
by measuring the shape of cyclotron orbits.\cite{Santos}
Another possible approach is to observe the ultrasonic absorption spectrum of
the sample.
The energy of an ultrasonic wave will be absorbed when it is in resonance with
the hyperorbits.
Similar method has been used to detect the existence of composite fermions
in half-filled quantum Hall systems.\cite{CF}

\subsection{Quantization of hyperorbits}

In a previous paper, we have derived the quantization condition using
Lagrangian formulation
combined with path integral method.\cite{Changprl} Here, it will be rederived
using a
slightly different approach.
Substituting ${\bf \dot r}={\bf \dot k}\times {\hat z}(\hbar/e\delta B)$ into
the
Lagrangian in Eq.~(2.17), we will obtain an
effective Lagrangian for the quasimomentum ${\bf k}$,
\begin{equation}
L({\bf k},{\bf{\dot k}})=\frac{\hbar^2}{e\delta B}
(k_1{\dot k}_2-k_2{\dot k}_1)
+\hbar{\mbox{\boldmath{${\cal A}$}}}\cdot{{\bf\dot k}}-E({\bf k}).
\end{equation}
It can be easily shown that Eq.~(4.1) does follow from this $L$.
The generalized momentum for coordinate ${\bf k}$ is equal to
\begin{equation}
\mbox{\boldmath$\pi$}=\frac{\partial L}{\partial {\bf \dot k}}
=-\frac{\hbar^2}{2e\delta B}{\bf k}\times{\hat z}+\hbar \mbox{\boldmath${\cal
A}$}({\bf k}).
\end{equation}
This leads to the following effective Hamiltonian
\begin{equation}
H({\bf k}, \mbox{\boldmath$\pi$})\equiv \mbox{\boldmath$\pi$}\cdot {\bf \dot
k}-
L({\bf k},{\bf \dot k})=E({\bf k}).
\end{equation}
Notice that because $H$ does not depend on \mbox{\boldmath$\pi$}, the
coordinate ${\bf k}$ will be a constant
of motion and the dynamics is trivial. A Hamiltonian that gives correct
dynamics
will be given in Appendix A. Since it is not central to our derivation, we will
not discuss it here.

The quantization of hyperorbits is given by
$\oint \mbox{\boldmath$\pi$}\cdot d{{\bf k}}=(m+\gamma)h$,
where $m$ is a non-negative integer and $\gamma$ will be taken to be $1/2$.
It leads to area quantization in ${\bf k}$-space
\begin{equation}
\frac{1}{2}\oint_{C_m} ({\bf k}\times d{\bf k})\cdot{\hat
z}=2\pi\left(m+\frac{1}{2}
-\frac{\Gamma(C_m)}{2\pi}\right)\frac{e\delta B}{\hbar},
\label{Qz}
\end{equation}
where
\begin{equation}
\Gamma(C_m)= \oint_{C_m}{\mbox{\boldmath{${\cal A}$}}}\cdot d{\bf k}
\end{equation}
is the Berry phase for orbit $C_m$. The orientation of $C_m$ is chosen such
that the
sign of the area on the left hand side of Eq.~(4.5) equals the sign of $\delta
B$.

The total number of hyperorbits in a MBZ is determined by requiring the area of
the outer-most
orbit be smaller than the area of a MBZ.
Assume the flux before perturbation is $B_0 a^2=p/q$,
then the number of hyperorbits in a MBZ is equal to
$\left| 1/(q\delta \phi)+\sigma \right|$,\cite{Changprl}
where $\delta \phi\equiv\delta B a^2 e/h$, and $\sigma$ is the Hall
conductivity of the
parent band. These hyperorbits are the lowest order approximation to  the
split energy subbands. They will be broadened by tunnelings between degenerate
orbits.
Since the MBZ is $q$-fold degenerate, the above number has to be divided by $q$
to get the actual number of daughter bands,
\begin{equation}
D=\frac{|1/(q\delta \phi)+\sigma|}{q}.
\end{equation}
This formula is essential in understanding the splitting pattern of the
Hofstadter spectrum.

The Hall conductivity for a subband can be calculated in the following way:
In the presence of both $\bf E$ and $\delta{\bf B}$,
the velocity of a magnetic Bloch electron consists of two parts,\cite{Mermin}
\begin{equation}
{\bf{\dot r}}=\frac{\hbar}{e\delta B}{\bf{\dot k}}\times {\hat z}+
\frac{{\bf E}\times{\hat z}}{\delta B}.
\end{equation}
The first term is the velocity of revolution, and the second term is the
velocity of drifting
along ${\bf E}\times {\bf B}$ direction.
The current density for a filled subband is
\begin{equation}
{\bf J}=-\int d^2{\bf k} \frac{\hbar}{\delta B}{\bf{\dot k}}\times {\hat z}-e
\int d^2{\bf k} \frac{{\bf E}\times{\hat z}}{\delta B}.
\end{equation}
The first integral is zero for a closed orbit. Therefore, the Hall conductivity
is obtained from the drifting term, which leads to $\sigma_{yx} ({\rm or\
simply}\
\sigma)=\rho e/
\delta B$, where $\rho$ is the electron density per unit area. This will be
used in the
next section to determine the Hall conductivities for subbands in the
Hofstadter spectrum.

\section{The Hofstadter spectrum}

\subsection{Hierarchical structure of the spectrum}

The discussion in the preceding section presumes that we know
$\delta B$. But this is not apparent if the magnetic field $B$ is homogeneous.
In this
case, we can still divide it into two parts, but
where is the dividing point between $B_0$ and $\delta B$?
A natural way of dividing $B$ (or $\phi$) is to write it as a continuous
fraction,
\begin{equation}
\phi={1\over\displaystyle f_1+
{\strut 1\over\displaystyle f_2+{\strut 1\over\displaystyle
{\strut {f_3+\cdots}}}}},
\end{equation}
and truncate it according to the accuracy we need.
The $r$-th order approximation of $\phi$ will be written as
$p_r/q_r$.\cite{Khinchin}  For example,
if $\phi=1/(2+\sqrt{2})$, we will have
$p_1/q_1=1/3,\ p_2/q_2=2/7,\ p_3/q_3=5/17 \cdots$ etc,
which are truncations of
\begin{equation}
\frac{1}{2+\sqrt{2}}={1\over\displaystyle 3+
{\strut 1\over\displaystyle 2+{\strut 1\over\displaystyle
{\strut {2+\cdots}}}}}.
\end{equation}

The $q_r$'s satisfy the following recursion relation
\begin{equation}
q_{r+1}=f_{r+1}q_r+q_{r-1},
\label{recursion}
\end{equation}
which relates the number of subbands at
neighboring orders. There is also a relation between $p$'s and $q$'s,
\begin{equation}
p_{r+1}q_{r}-p_{r}q_{r+1}=(-1)^{r}.
\end{equation}
It follows that the extra magnetic flux between the $r$-th order and the
$(r+1)$-th order
truncation is
\begin{equation}
\delta
\phi_r=\frac{p_{r+1}}{q_{r+1}}-\frac{p_{r}}{q_{r}}=\frac{(-1)^r}{q_{r+1}q_r}.
\end{equation}
Since the sign of $\delta \phi_r$ alternates from one order to the next,
the direction of $\delta B$ also alternates.\cite{even}

Notice that for a chosen fraction $p_r/q_r$, the size of a MBZ is fixed.
Without perturbation from the part that is truncated away,
a wave packet will move on a straight line.
The trajectory is curved because of $\delta \phi_r$. Higher the order
approximation we
use, smaller the $\delta \phi$ gets and larger the
radius of the hyperorbit becomes. In the ideal case, without any complications
due to disorder, thermal broadening $\cdots$ etc,
we expect there will be a hierarchical structure of hyperorbits due to
different orders of approximation. This structure finds its
correspondence in the hierarchical structure of the Hofstadter
spectrum.\cite{lit}
(Fig.~2)
\vspace{1 cm}
\begin{figure}
\epsfxsize=5truein
\hskip 0.0truein
\epsffile{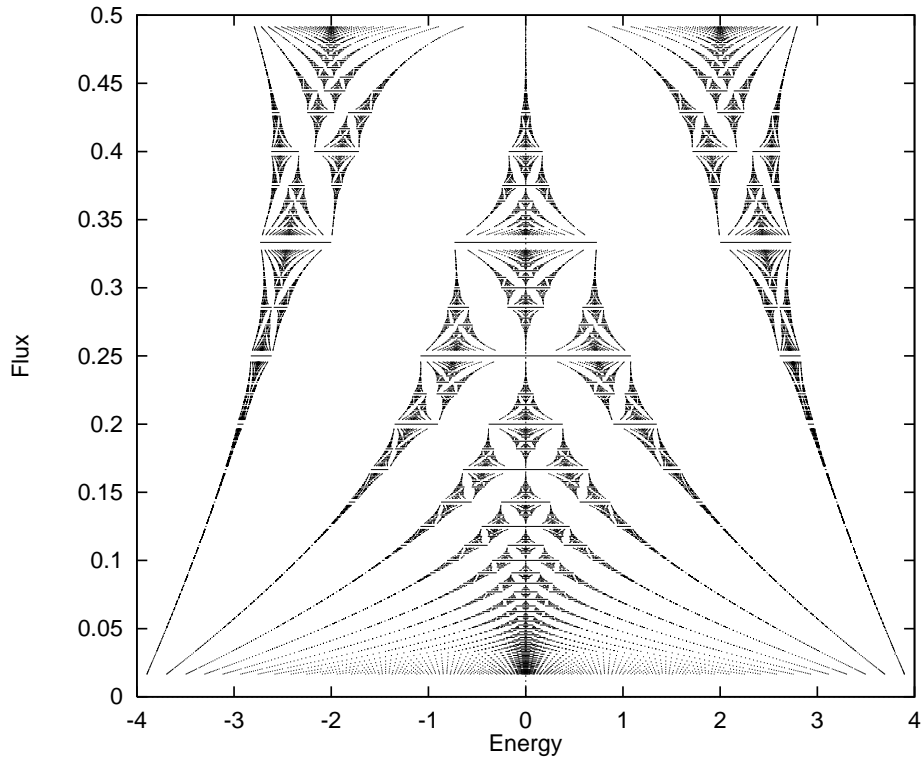}
\vspace{.8 cm}
\caption{Hofstadter spectrum for a square lattice generated from a
tight-binding model
($q\leq 60$). Only the $\phi\leq 1/2$ part is shown since the spectrum is
symmetric with respect to $\phi=1/2$.}
\end{figure}

\subsection{Distribution of Hall conductivities and splitting of energy bands}

A magnetic Bloch band carries quantized Hall current, and this current will
redistribute among
daughter bands in such a way that the total Hall current for subbands
equals the original current.\cite{Avron}
We will call this the ``sum rule". The current distribution among
subbands, which is also quantized in each subband,
was obtained by Thouless {\it et al}.\cite{TKNdN}  In their famous paper,
they found that the subband Hall conductivities are the integer-valued
solutions of
the Diophantine equation.
Here we show that semiclassical dynamics offers an alternative and very
heuristic solution to
this problem. The Hall conductivities calculated will be used in
Eq.~(4.7) to calculate the number of magnetic subbands after splitting.

The general expression for the Hall conductivity of a ``closed" subband is
$\sigma=\rho e/\delta B$ (see Eq.~(4.9) and below). Therefore, $\sigma$ can be
determined
if $\rho$ and $\delta B$ are known. Consider
a subband at the $r$-th order of splitting. Since all subbands at this level
share the same number
of states, each subband will have  $\rho_r=\rho_0/q_r$,
where $\rho_0$ is the density of states for the original Bloch band. The
perturbation field for a
subband at this level is $\delta B_r=h\delta \phi_{r-1}/(ea^2)$.
Therefore, we have
\begin{equation}
\sigma^{\rm close}_r=\frac{e\rho_r}{\delta B_r}=(-1)^{r-1}q_{r-1},\ ({\rm in\
units\ of}\ e^2/h).
\end{equation}
(Since $\sigma_1=1$ for a closed subband at the first level, $q_0$ will be set
to one .)
We have to emphasize that, Eq.~(5.6) is valid for {\it every closed subbands}
at the
$r$-th order. Combining Eq.~(4.7) with (5.6), we can determine the number of
daughter bands being split from a $r$-th order parent band, which is
\begin{eqnarray}
{\cal D}^{\rm close}_r&=&\frac{|(-1)^r q_{r+1}+\sigma_r^{\rm close}|}{q_r}\cr
&=&f_{r+1}.
\end{eqnarray}

\begin{figure}
\epsfxsize=3truein
\hskip 1.0truein \epsffile{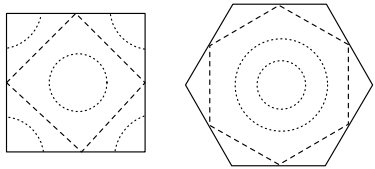}
\caption{Schematic plots of the constant energy contours in the
reduced MBZ of
a square lattice and a triangular lattice. Dashed lines are the open
orbits.}
\end{figure}
The Hall conductivity for an open subband is more difficult to obtain. It
requires the knowledge of the exact ${\bf k}$-trajectory  to figure out the
first integral in
Eq.~(4.9). However, for a square or a  triangular lattice, there is an easier
way of calculating
it. This is so because, there is only
one open orbit in every MBZ for either lattice (see Fig.~3).
Therefore, there is only one open
daughter band for every parent band. Its Hall conductivity can be figured out
by using the
``sum rule":
\begin{equation}
\sigma_{\rm parent}=\sum \sigma_{\rm daughter}.
\end{equation}
For example, at the first order, we have
$\sigma_1^{\rm open}=\sigma_0-(f_1-1)\sigma_1^{\rm close}=-(f_1-1)$,
where $\sigma_0=0$ since it is a Bloch band at the very beginning.
For $r\ge 2$, the Hall conductivity of an open daughter band at the
$r$-th order is
\begin{eqnarray}
\sigma^{\rm open}_r
&=&\sigma_{r-1}^{\rm close}-(f_r-1)\sigma_r^{\rm close}\cr
&=&(-1)^{r-2}q_{r-2}-(f_r-1)(-1)^{r-1} q_{r-1}\cr
&=&(-1)^{r-1}q_{r-1}+(-1)^rq_r,
\end{eqnarray}
where we have assumed that its parent is a 'closed' band.
Using Eq.~(4.7), we see that this open band (now as a parent) will split into
\begin{equation}
{\cal D}^{\rm open}_r=f_{r+1}+1
\end{equation}
subbands under perturbation. It can be
shown that the same result as Eq.~(5.9) is obtained if its parent is an
'open' band (with $f_r+1$ daughters). This checks the consistency of this
calculation.
It is clear that the extra splitting of one subband
from each open parent band (there are $q_{r-1}$ of them) accounts for the extra
$q_{r-1}$ in the recursion relation Eq.~(\ref{recursion})

We give one example to demonstrate the use of these rules.
Consider a square lattice with $\phi=1/(2+\sqrt{2})$.
Because $\sigma_1^{\rm close}=1$, and $\sigma_1^{\rm open}=-2$,
the distribution of $\sigma$'s for the three subbands
at the first order is
\begin{equation}
\mbox{\boldmath $\sigma_{\rm 1}$}=(1,-2,1).
\end{equation}
where we have put $\sigma_1^{\rm open}$ in the middle since for a square
lattice the open subband is located at the center of a parent band (Fig.~3).

These three bands will be split into seven subbands due to the extra flux
$\delta \phi_1=p_2/q_2-p_1/q_1=2/7-1/3$.
Since $f_2=2$, the pattern of splitting will be, according to
Eqs.~(5.7) and (5.10),
\begin{equation}
\mbox{\boldmath ${\cal D}_{\rm 1}$}=(2,3,2).
\end{equation}
Furthermore, since $\sigma^{\rm close}_2=-3$ and $\sigma^{\rm open}_2=4$
according
to Eqs.~(5.6) and (5.9), the Hall conductivity distribution is \cite{edge}
\begin{equation}
\mbox{\boldmath $\sigma_{\rm 2}$}=(-3,4,-3,4,-3,4,-3).
\end{equation}
Consequently, we have
\begin{equation}
\mbox{\boldmath ${\cal D}_{\rm 2}$}=(2,3,2,3,2,3,2).
\end{equation}
The Hall conductivities we just obtained are the same as those
derived from the Diophantine equation.
Actual pattern of splitting is shown in Fig.~4 for comparison.
The distribution in Fig.~4 for the left (or right) five subbands in Eq.~(5.14)
appears to be $(2,1,2)$, instead of $(2,3)$ (or $(3,2)$). However, closer
examination
reveals that the left (right)  three subbands actually come from the same
parent.
In fact, slight asymmetry in the distribution is inevitable
because when $\phi$ is changed by a small amount, an electron state cannot
suddenly
jump out of the band edge to the middle of a gap.

Eqs.~(5.6)--(5.10) also apply to a triangular lattice. The only difference is
that
$\sigma_{\rm open}$ no longer locates at the center of a parent band.
Given the same $\phi=1/(2+\sqrt{2})$, we now have
\begin{eqnarray}
\mbox{\boldmath $\sigma_{\rm 1}$}&=&(1,1,-2) \cr
\mbox{\boldmath ${\cal D}_{\rm 1}$}&=&(2,2,3)\cr
\mbox{\boldmath $\sigma_{\rm 2}$}&=&(4,-3,-3,4,4,-3,-3)\cr
\mbox{\boldmath ${\cal D}_{\rm 2}$}&=&(3,2,2,3,3,2,2).
\label{list}
\end{eqnarray}
This again conforms with the actual spectrum and the solutions of the
Diophantine
equation with subsidiary constraints suitable for a triangular
lattice.\cite{Diophantine}
\begin{figure}
\epsfxsize=3truein
\hskip 1.0truein \epsffile{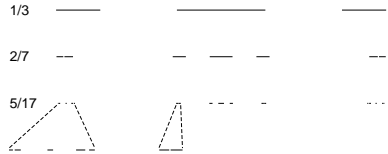}
\caption{Pattern of energy splitting for $\phi$ in Eq.~(5.2).
Different orders of approximation give
3, 7, and 17 subbands respectively.}
\end{figure}

\section{Calculations of energy spectrum, curvature, magnetic moment, and
cyclotron energy}

In this section, we give detailed calculations of ${\cal E}_n({\bf k})$,
$\Omega_n({\bf k})$,
and $L_n({\bf k})$.
They are used in calculating the cyclotron energies using the
quantization formula Eq.~(\ref{Qz}). The cyclotron energies will be used to
estimate
the subband energies in the Hofstadter spectrum.
In doing so, we not only presume the one-band approximation (on which Fig.~2 is
based),
but also neglect the inter-orbit transitions that broaden (and may slightly
shift)
the energy levels. The latter approximation
leads to negligible error if the bandwidths under
consideration are very small. Similar calculations have been done by
Wilkinson, and his results have been very successful.\cite{Wlist} However, we
believe that
the approach proposed here is conceptually simpler, and is easier to
generalize to other types of lattices.

\subsection{Calculation of energy spectrum for parent bands}

The following calculation is based on the tight-binding model.
We will only give a very brief explanation of this approach.
For more details, we request readers to refer to Ref.~(29).
In the tight-binding approximation, a Bloch state is expanded as (for
$\phi=p/q$)
\begin{equation}
\Psi({\bf k})=\sum_{l=1}^q a_l({\bf k}) \psi_l({\bf k}),
\end{equation}
where ${\bf k}$ is restricted to a reduced MBZ.
The basis $\psi_l({\bf k})$ is defined to be $\psi(k_1+2\pi \phi l,k_2)$, where
$\psi({\bf k})$ is a Bloch state before the magnetic perturbation.
A tight-binding Hamiltonian in the absence of $\delta B$,
when being expressed on the basis of $\psi_l$, is a
$q\times q$ matrix (the lattice constant $a$ is set to one.)
\begin{equation}
H_{l',l}=-\left(\begin{array}{cccccc}
\cos(k_1+2\pi/q)& e^{ik_2}               & 0  & \cdot   & 0 & e^{-ik_2}  \\
e^{ik_2}               &\cos(k_1+4\pi/q) & e^{ik_2}      & \ddots  &\ddots  & 0
\\
0          & e^{ik_2}               &\ddots    & \ddots       & \ddots  &\cdot
   \\
\cdot   & \ddots  & \ddots  & \ddots   & e^{ik_2}       & 0\\
0         & \ddots   & \ddots  & e^{ik_2} & \cos(k_1+(q-1)\frac{2\pi}{q})  &
e^{ik_2} \\
e^{-ik_2}    &  0   & \cdot    &0            & e^{ik_2}   & \cos(k_1)  \\
\end{array}\right).
\label{matrix}
\end{equation}
${\cal E}_n({\bf k})$ and $a^n_l({\bf k})$ are nothing but the eigenvalues and
eigenvectors of this matrix.
For example, if $p/q=1/3$, then a straightforward calculation shows that the
${\cal E}_n({\bf k})$'s are solutions of the following characteristic equation,
\begin{equation}
-{\cal E}^3+6{\cal E}=2\left[\cos(3k_1)+\cos(3k_2)\right].
\label{eigen}
\end{equation}
There are three roots for each ${\bf k}$, and variation of ${\bf k}$ over the
MBZ leads to the three energy
bands for $\phi=1/3$ (see Fig.~2). It is not difficult to see that
the band edges are located at (from high to low) ${\cal E}_1({\bf g}), {\cal
E}_1({\bf 0}),
{\cal E}_2({\bf 0}), {\cal E}_2({\bf g}), {\cal E}_3({\bf g})$, and ${\cal
E}_3({\bf 0})$,
where ${\bf g}= (\pi/3,\pi/3)$.

\subsection{Calculation of Berry curvature}

To calculate the Berry curvature in Eq.~(\ref{master}), we need to know the
eigenvectors of
$H_{l'l}$. Before doing that, we will try to rewrite Eq.~(\ref{master})
in a form suitable for the tight-binding calculation.
First, we insert a complete state $\sum_{n'} |u_{n'}\rangle \langle u_{n'}|$
inside the
dot products that appear in
Eq.~(\ref{master}). Since we are using the one-band approximation, $n'$
only runs through subbands in the same parent
band, and
\begin{equation}
\Omega_n({\bf k})= i\sum_{n'=1}^q\ '
\left[ {\langle \frac{\partial u_n}{\partial k_1}|u_{n'}\rangle \langle u_{n'}|
\frac{\partial u_n}{\partial k_2}\rangle- {\rm c.c.}} \right],
\end{equation}
where we have dropped a term with $n'=n$ since
$\langle u_n|\partial /\partial {\bf k}|u_n\rangle$ is purely imaginary and
does not contribute
to the curvature. With the help of the identity
\begin{equation}
\langle u_{n'}|\frac{\partial }{\partial {\bf k}}|u_n\rangle=
\frac{\langle u_{n'}|\frac{\partial \tilde H}{\partial {\bf k}}| u_n\rangle}
{{\cal E}_{n'}-{\cal E}_n},
\end{equation}
where $\tilde H \equiv e^{-i{\bf k}\cdot {\bf r}} H e^{i{\bf k}\cdot {\bf r}}$
(this and the following $H$'s are the unperturbed Hamiltonian in Eq.~(1.3);
the subscript '0' is dropped for brevity),
we can rewrite Eq.~(6.4) in the form
\begin{equation}
\Omega_n({\bf k})= i\sum_{n'}\ ' \left[
\frac{\langle u_n|\frac{\partial \tilde H}{\partial k_1}|u_{n'}\rangle
\langle n_{n'}|\frac{\partial \tilde H}{\partial k_2}| u_n\rangle}
{({\cal E}_{n'}-{\cal E}_n)^2} -{\rm c.c.}\right].
\label{TBO}
\end{equation}
Expanding $u_n$ by $\tilde \psi_{l{\bf k}}({\bf r})$, which is defined to be
$e^{-i{\bf k}\cdot{\bf r}}\psi_{l{\bf k}}({\bf r})$, we have
\begin{eqnarray}
\langle u_{n'}|\frac{\partial \tilde H}{\partial {\bf k}}| u_n\rangle
&=&\sum_{l',l} a^{n'*}_{l'} a^n_l\langle \tilde\psi_{l'}|\frac{\partial \tilde
H}
{\partial {\bf k}}|\tilde\psi_l\rangle \cr
&=& \sum_{l',l} a^{n'*}_{l'} a^n_l \frac{\partial}{\partial {\bf k}}\langle
\tilde\psi_{l'}| \tilde H |\tilde\psi_l\rangle \cr
&+& ({\cal E}_n-{\cal E}_{n'})\sum_{l',l}
a^{n'*}_{l'} a^n_l \langle \tilde\psi_{l'} | \frac{\partial
\tilde\psi_l}{\partial {\bf k}} \rangle.
\end{eqnarray}
The inner product $\langle \tilde\psi_{l'} | \partial
\tilde\psi_l/\partial {\bf k} \rangle$ is zero
for a Bloch state in the absence of a magnetic field. Therefore,
\begin{equation}
\langle u_{n'}|\frac{\partial \tilde H}{\partial {\bf k}}| u_n\rangle
=\sum_l a^{n'*}_l a^n_l\frac{\partial H_{l',l}}{\partial {\bf k}}.
\label{core}
\end{equation}
Beyond this stage, the calculation is straightforward since we only need to
calculate
$a^n_l$ from Eq.~(\ref{matrix}) and combine Eqs.~(\ref{TBO}) and (6.8) to
obtain
$\Omega_n({\bf k})$.

Again we choose a simple fraction $p/q=1/3$ and calculate
the Berry curvature distributions for the three magnetic subbands.
The result is shown in Fig.~5, in which the range of ${\bf k}$ vector is one
reduced MBZ (the basic unit of repetition).
Note that the curvature tends to concentrate on four inner band edges
because the electron states near inner gaps
are changed the most from the original Bloch states that have zero Berry
curvature.
The curvatures from the three bands cancel locally, that is
\begin{equation}
\sum_{n=1}^3 \Omega_n({\bf k})=0,\ \ \forall \ {\bf k}.
\end{equation}
This is in general true for any $q$ and can be easily proved from
Eq.~(\ref{TBO}).
Finally, integration of
$\Omega_n({\bf k})$ over a MBZ divided by $2\pi$ gives us integers $(1,-2,1)$.
These are indeed the Chern numbers we expected (see Eq.~(5.11)).
\begin{figure}
\epsfxsize=3truein
\hskip 1.0truein \epsffile{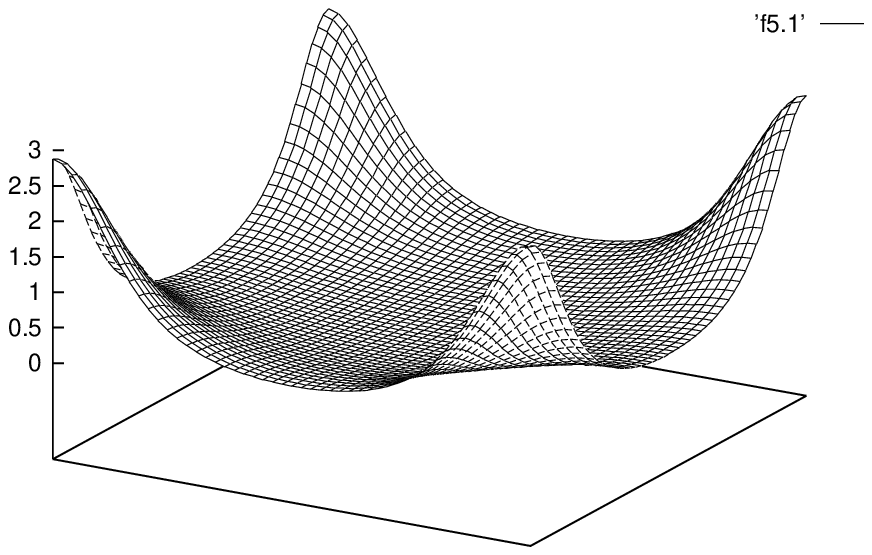}
\end{figure}
\begin{figure}
\epsfxsize=3truein
\hskip 1.0truein \epsffile{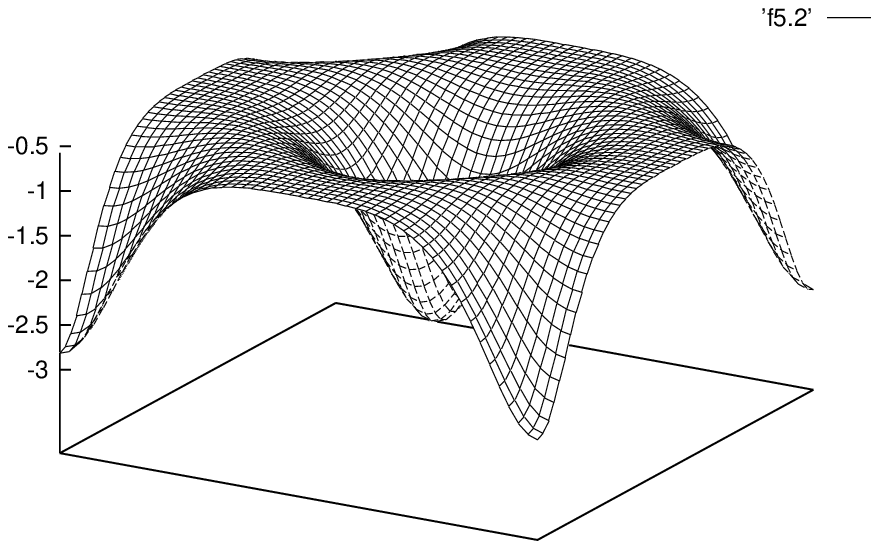}
\end{figure}
\begin{figure}
\epsfxsize=3truein
\hskip 1.0truein \epsffile{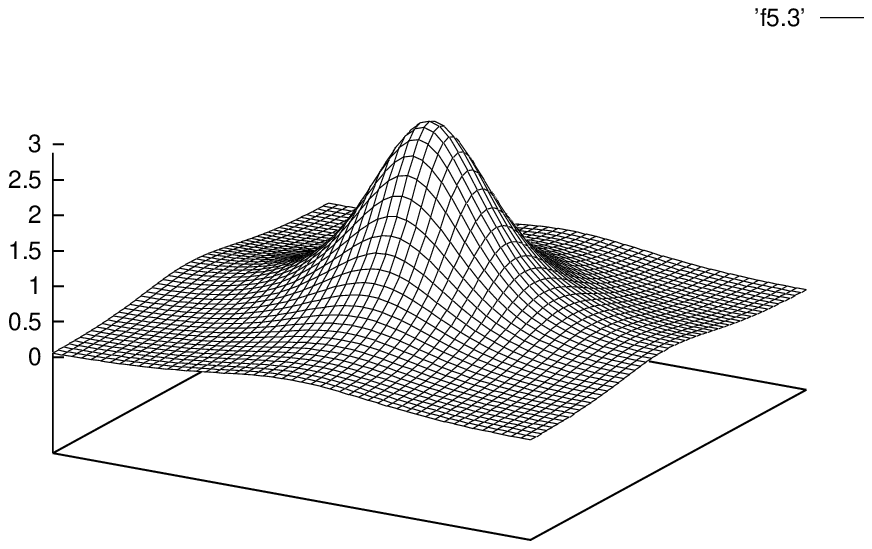}
\caption{Distributions of Berry curvature $\Omega_n({\bf k})$ (in
units of
$e^2/h$). $\Omega_1({\bf k})$ is equal to $\Omega_3({\bf k})$ shifted by
$(\pi/3,\pi/3)$}
\end{figure}

\subsection{Calculation of magnetic moment}

The self-rotating angular momentum of the wave packet $\langle {\bf
L}\rangle_{n{\bf k}}$ in
Eq.~(2.13) can be written in a from that is more
tractable for calculation:
\begin{equation}
L_n({\bf k})= i\frac{m}{\hbar}\left[\langle\frac{\partial u_n}{\partial k_1}|
\tilde H-{\cal E}_n|\frac{\partial u_n}{\partial k_2}\rangle-{\rm c.c.}\right].
\label{Bellissard}
\end{equation}
Derivation of this formula is given in Appendix B.
Eq.~(\ref{Bellissard}) can also be rewritten as
\begin{equation}
L_n({\bf k})=i\frac{m}{\hbar}\sum_{n'}\ ' \left[
\frac{\langle u_n|\frac{\partial \tilde H}{\partial k_1}|u_{n'}\rangle
\langle n_{n'}|\frac{\partial \tilde H}{\partial k_2}| u_n\rangle}
{{\cal E}_{n'}-{\cal E}_n} -{\rm c.c.}\right].
\label{TBM}
\end{equation}
Derivation of Eq.~(\ref{TBM}) is very similar to the derivation of
Eq.~(\ref{TBO}). The only
change is that the extra factor of $\tilde H-{\cal E}_n$ in the numerator
cancels a ${\cal E}_{n'}-{\cal E}_n$ in the denominator.

$L_n({\bf k})$ can be readily calculated by combining Eq.~(\ref{core}) with
Eq.~(\ref{TBM}).
The result is shown in Fig.~6 (again for $\phi=1/3$).
Similar to the distribution of $\Omega_n({\bf k})$, $L_n({\bf k})$
also has peaks near inner band edges. The total magnetizations from all three
bands
cancel each other. However, unlike the curvature, they do not cancel locally at
each ${\bf k}$ point.

We remark that this magnetization energy first appeared in a paper by
Kohn\cite{history}. Their objective was to study the effective
Hamiltonian for
Bloch electrons in a weak electromagnetic field.
This term is in general zero for Bloch bands, but can be nonzero in the
presence of
spin-orbit interaction. In the latter case, it contributes an extra g-factor to
Bloch
electrons.\cite{g} An expression that is the same as the right hand side of
Eq.~(\ref{Bellissard})
has also been obtained by Rammal and Bellissard.\cite{Bellissard}
Without the $m/\hbar$ factor (and apart from a factor of two),
it is called the Rammal-Wilkinson form.
\begin{figure}
\epsfxsize=3truein
\hskip 1.0truein \epsffile{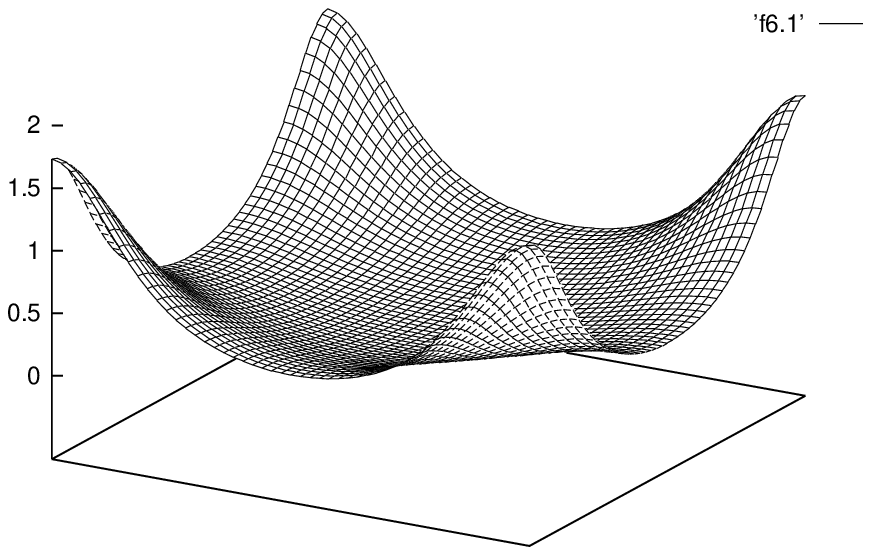}
\end{figure}
\begin{figure}
\epsfxsize=3truein
\hskip 1.0truein \epsffile{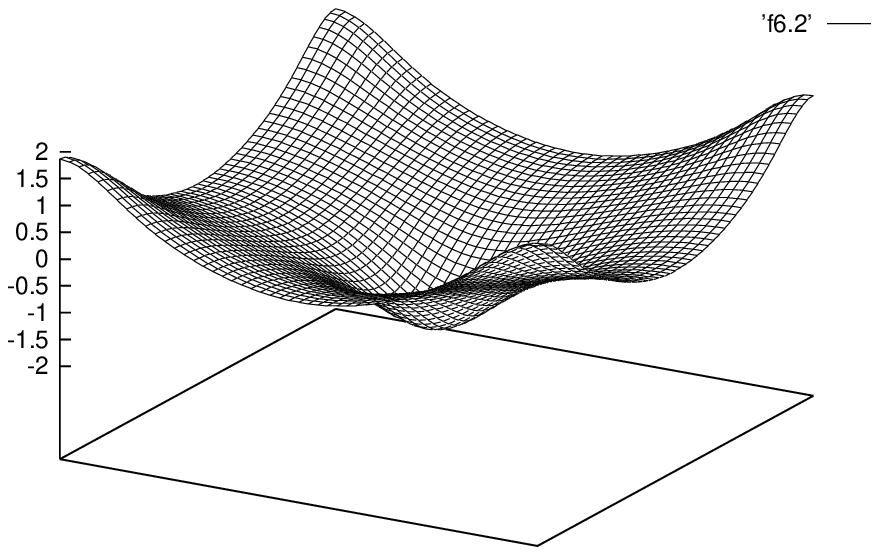}
\end{figure}
\begin{figure}
\epsfxsize=3truein
\hskip 1.0truein \epsffile{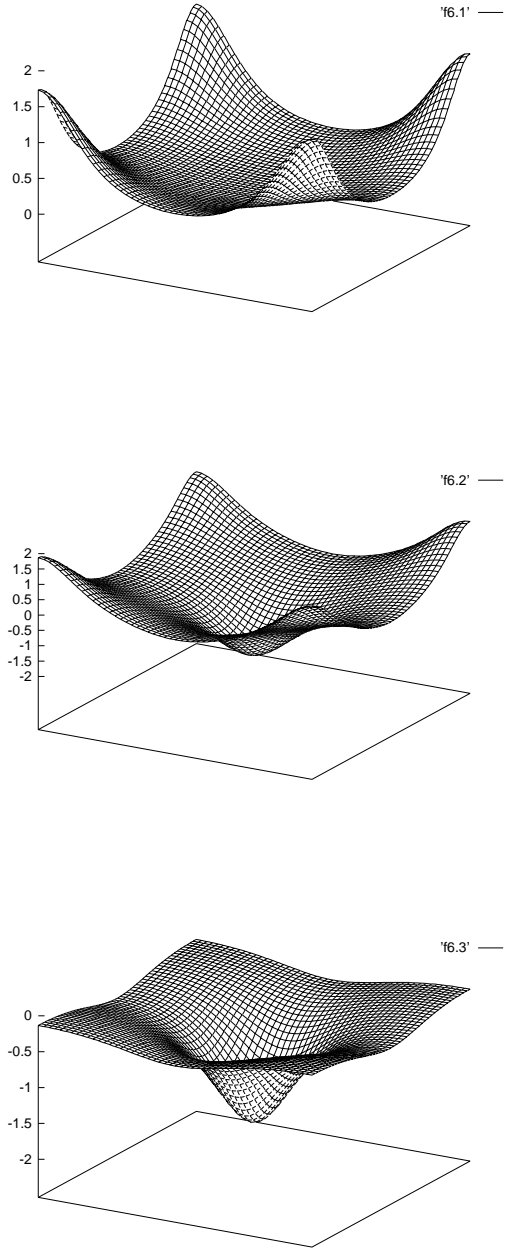}
\caption{Distributions of angular momentum $L_n({\bf k})$ (in
units of $2m/\hbar$).}
\end{figure}

\subsection{Calculation of cyclotron energy}

After obtaining ${\cal E}_n({\bf k})$, $\Omega_n({\bf k})$, and $L_n({\bf k})$,
we can determine the cyclotron energies according to
the quantization formula Eq.~(\ref{Qz}).
In the following example, we add $\delta \phi=-1/201$ to $\phi=1/3$.
This gives $\phi'=22/67=1/(3+1/22)$. According to the simple rules derived in
previous section, the original magnetic bands are
expected to split to 22, 23, and 22 subbands respectively.
In Table II, we compare the exact spectrum with the quantized cyclotron
energies. Only
part of the subbands from the parent band in the middle are shown.
We show two numbers (for two band edges) in the first row
when the bandwidth for $E_{\rm Hofst}$ is larger than $10^{-4}$.
The $E_{\rm cyclo}$'s in the second row are obtained by fine-tuning the
path $C_m$ in Eq.~(\ref{Qz}), with uncertainty on the order of $10^{-4}$.
It can be seen that the match between $E_{\rm cyclo}$ and $E_{\rm Hofst}$ is
quite satisfying.
We have done calculations for subbands from other parent bands, and they also
show
similar accuracy.

Notice that the energy levels from fractions like $\phi=1/f$ are broadened
cyclotron levels in an ordinary Bloch band. They do not split from a magnetic
parent band.
In this case $\Omega_n({\bf k})$ and $L_n({\bf k})$ are zero, and
Eq.~(\ref{Qz}) reduces to the usual Onsager quantization formula.
Numerical result based on this simplified formula for the cyclotron energies
also agrees very well with the positions of subbands in Fig.~2.

\section{Summary}

Electron states in a lattice subject to a homogeneous magnetic field satisfy
magnetic translation symmetry and have band-like energy spectrum
similar to the usual Bloch band. However, to our knowledge,
the semiclassical dynamics of magnetic Bloch electrons has never been an
explicit subject.
One reason is that the observation of band-splitting remains an experimental
challenge to date;
the other reason might be due to the fact that magnetic Bloch bands, unlike
Bloch bands,
can be changed easily by varying an external magnetic field. However, our study
has shown
that the inquiry of magnetic Bloch bands can be very rewarding in itself.
Major findings in this paper are summarized below:

The Berry curvature of magnetic bands plays a crucial role in the
dynamics. It gives electrons an extra velocity in the direction of ${\bf
E}\times{\bf B}$, and
this term directly relates to the quantization of Hall conductivity.
This semiclassical dynamics, combined with the Boltzmann equation,
is used to study electron transport in a DC or AC electric field.

In the presence of $\delta B$, the energy dispersion $E({\bf k})$
is shifted from the usual band energy because of the non-zero magnetic moment.
Similar to usual Bloch electrons,
magnetic Bloch electrons execute cyclotron motion on the constant energy
surface of $E({\bf k})$. However, the quantization condition for cyclotron
orbits
has to be modified from the usual Onsager condition because of
the Berry phase. Based on this modified formula, we obtain a simple rule
that calculates the number
of daughter bands for every parent band in the Hofstadter spectrum.
Furthermore, a fairly heuristic explanation for the distribution of Hall
conductivities is given using the picture of cyclotron orbit drifting.

These quantized orbits are closely related to the energy bands in the
Hofstadter spectrum.
We give detailed numerical calculations for $\Omega_n({\bf k})$ and $L_n({\bf
k})$
based on the tight-binding model for the case of $\phi=1/3$.
They are used in the calculation of cyclotron energies using the quantization
condition,
and the result is in very good agreement with the actual spectrum.
This shows that the complex pattern of the
Hofstadter spectrum is nothing more than the broadened cyclotron energy
spectrum in
magnetic Bloch bands.

\acknowledgments
The authors wish to thank A.~Barr, J.~Bellissard, F.~Claro, E.~Demircan,
G.~Georgakis,
M.~Kohmoto, W.~Kohn, M.~Marder, G.~Sundaram, and M.~Wilkinson
for many helpful discussions. This work is supported by the R.~A.~Welch
foundation.

\appendix

\section{}

Note that there is no ${\bf\dot k}$ dependence in \mbox{\boldmath$\pi$}
(because there is no kinetic energy term ${\bf\dot k}^2$ in $L({\bf k},\dot{\bf
k})$). In this case,
it is more pertinent to treat Eq.~(4.3) as a constraint on the variables ${\bf
k}$ and
\mbox{\boldmath$\pi$},
\begin{equation}
\mbox{\boldmath$\theta$}({\bf k},\mbox{\boldmath$\pi$})
\equiv \mbox{\boldmath$\pi$}+\frac{\hbar^2}{2e\delta B}
{\bf k}\times{\hat z}-\hbar \mbox{\boldmath${\cal A}$}({\bf k})\simeq 0.
\end{equation}
Strictly speaking, this constraint cannot be used before obtaining the
equations of motion,
therefore we use $\simeq$ to
distinguish it from a real identity.\cite{Dirac}
A general Hamiltonian for a system with constraints is given by,
\begin{equation}
H^*({\bf k},\mbox{\boldmath$\pi$})=H({\bf
k},\mbox{\boldmath$\pi$})+{\mbox{\boldmath$\lambda$}}
\cdot {\rm \mbox{\boldmath$\theta$}}=
E({\bf k})+{\mbox{\boldmath$\lambda$}}
\cdot {\rm \mbox{\boldmath$\theta$}},
\end{equation}
where $\mbox{\boldmath$\lambda$}=(\lambda_1,\lambda_2)$ are arbitrary functions
of
${\bf k}$ and \mbox{\boldmath$\pi$}. The dynamical equations for this new
Hamiltonian are
\begin{eqnarray}
{\bf\dot k} &=& \frac{\partial H^*}{\partial \mbox{\boldmath$\pi$}}
=\mbox{\boldmath$\lambda$}\cr
{\bf\dot{\mbox{\boldmath$\pi$}}}
&=& -\frac{\partial H^*}{\partial {\bf k}}=
-\frac{\partial E}{\partial {\bf k}}+
\frac{\hbar^2}{2e\delta B}\mbox{\boldmath$\lambda$}\times{\hat z}
+\hbar\sum_i \lambda_i\frac{\partial  {\cal A}_i}{\partial {\bf k}},
\end{eqnarray}
where we have discarded a term $\partial \mbox{\boldmath$\lambda$}/\partial{\bf
k}
\cdot{\mbox{\boldmath$\theta$}}=0$.
According to Eq.~(4.3), we should also have
\begin{equation}
{\bf\dot {\mbox{\boldmath $\pi$}}}
=-\frac{\hbar^2}{2e\delta B}{\bf \dot k}\times {\hat z}
+\hbar\sum_i {\dot k}_i\frac{\partial {\mbox{\boldmath{${\cal A}$}}}}{\partial
k_i}.
\end{equation}
Equating (A3) to (A4), and replacing {\mbox{\boldmath$\lambda$}} by
${\bf\dot k}$, we will get Eq.~(4.1).

\section{}

We will rewrite the angular momentum of a wave packet in Eq.~(2.13) in terms of
magnetic Bloch
functions. By defining $\tilde w({\bf k})=e^{i{\bf k}\cdot {\bf r}_c} w({\bf
k})$, we have
\widetext
\begin{eqnarray}
{\bf L}_n({\bf k}_c)&=& \int d^2 {\bf k}'\int d^2 {\bf k} \ w^*({\bf k}')
w({\bf k})
\langle \Psi_n({\bf k}')|({\bf r}-{\bf r}_c)\times {\bf P}|\Psi_n({\bf
k})\rangle \cr
&=& \int d^2 {\bf k}'\int d^2 {\bf k}\  \tilde w^*({\bf k}') \tilde w({\bf k})
\langle u_n({\bf k}')|e^{i({\bf k}-{\bf k}')\cdot({\bf r}-{\bf r}_c)}({\bf
r}-{\bf r}_c)\times {\bf \tilde P}({\bf k})|
u_n({\bf k})\rangle,
\end{eqnarray}
\narrowtext
where ${\bf \tilde P}$ is the momentum operator on the $|u_n\rangle$ basis.
Since
\begin{equation}
{\bf \tilde P}({\bf k})|u_n({\bf k})\rangle=\sum_{n'} |u_{n'}({\bf k})\rangle
\langle u_{n'}({\bf k})|
{\bf \tilde P}({\bf k})|u_n({\bf k})\rangle,
\end{equation}
and
\begin{eqnarray}
&\ &\langle u_n ({\bf k}') | e^{i({\bf k}-{\bf k}')
\cdot({\bf r}-{\bf r}_c)} ({\bf r}-{\bf r}_c)|u_{n'}({\bf k})\rangle\cr
&=& i\delta_{n,n'}\frac{\partial}{\partial {\bf k}'} \delta({\bf k}-{\bf k}')
-i\langle \frac{\partial u_n}{\partial {\bf k}'}|e^{i({\bf k}-{\bf
k}')\cdot({\bf r}-{\bf r}_c)}
|u_{n'}\rangle\cr
&=& i\delta_{n,n'}\frac{\partial}{\partial {\bf k}'} \delta({\bf k}-{\bf k}')
-i\delta({\bf k}-{\bf k}')\langle \frac{\partial u_n}{\partial {\bf
k}'}|u_{n'}\rangle,
\end{eqnarray}
we have
\begin{eqnarray}
{\bf L}_n({\bf k}_c)=&-&i\int d^2 {\bf k}\left[
\frac{\partial}{\partial {\bf k}} \tilde w^*({\bf k})\right]\tilde w({\bf
k})\times
\langle {\bf P}\rangle_n\cr
&-&i\int d^2 {\bf k}~|\tilde w({\bf k})|^2
\langle \frac{\partial u_n}{\partial {\bf k}} | \times {\bf \tilde P}({\bf
k})|u_n\rangle.
\end{eqnarray}
Because ${\bf \tilde P}({\bf k})=(m/\hbar)\partial {\tilde H}/\partial {\bf
k}$,
the integrand of the second term can be written as
\widetext
\begin{eqnarray}
&\ &\frac{m}{\hbar}\langle \frac{\partial u_n}{\partial k_1}|\frac{\partial
\tilde H}
{\partial k_2} |u_n\rangle-(k_1\leftrightarrow k_2) \cr
&=& \frac{m}{\hbar}\frac{\partial}{\partial k_2}\langle \frac{\partial
u_n}{\partial k_1}|
\tilde H |u_n\rangle-\frac{m}{\hbar}\langle \frac{\partial u_n}{\partial
k_1}|\tilde  H|
\frac{\partial u_n}{\partial k_2}\rangle-(k_1\leftrightarrow k_2) \cr
&=&\frac{m}{\hbar}\langle \frac{\partial u_n}{\partial k_1}|
\frac{\partial u_n}{\partial k_2}\rangle{\cal E}_n+ \frac{m}{\hbar}
\langle\frac{\partial u_n}{\partial k_1} |u_n\rangle\frac{\partial {\cal
E}_n}{\partial k_2}
-\frac{m}{\hbar}\langle \frac{\partial u_n}{\partial k_1}|\tilde  H|
\frac{\partial u_n}{\partial k_2}\rangle
-(k_1\leftrightarrow k_2) \cr
&=&\frac{m}{\hbar}\langle \frac{\partial u_n}{\partial {\bf k}}|\times({\cal
E}_n-\tilde H)|
\frac{\partial u_n}{\partial {\bf k}}\rangle +
\langle \frac{\partial u_n}{\partial {\bf k}}|u_n\rangle \times\langle {\bf
P}\rangle_n.
\end{eqnarray}
\narrowtext
Therefore we have
\begin{eqnarray}
{\bf L}_n({\bf k}_c) &=& i\frac{m}{\hbar}\langle \frac{\partial u_n}{\partial
{\bf k}}|
\times(\tilde H-{\cal E}_n)|
\frac{\partial u_n}{\partial {\bf k}}\rangle|_{{\bf k}={\bf k}_c}\cr
&-& i\int d^2{\bf k}~|\tilde w({\bf k})|^2 \langle \frac{\partial u_n}{\partial
{\bf k}}| u_n\rangle\times
\langle {\bf P}\rangle_n \cr
&-& i\int d^2{\bf k} \left[ \frac{\partial}{\partial {\bf k}} \tilde w^*({\bf
k})\right]
\tilde w({\bf k})\times \langle {\bf P}\rangle_n.
\end{eqnarray}
The last two terms cancel because of Eq.~(\ref{condition}), and this leads to
Eq.~(\ref{Bellissard})
Notice that this result is independent of the way a wave packet is constructed
since there is no $w({\bf k})$-dependence in the new expression.

\widetext
\begin{table}
\caption{}
\begin{tabular}{lll}
& & \\
& Bloch band & Magnetic Bloch band ($(p/q)\phi_0$ per plaquette) \\ \hline
& &\\
Unperturbed Hamiltonian \ \ \ \ &
$H_0=\frac{1}{2m}\left( -i\hbar\frac{\partial}{\partial {\bf
r}}\right)^2+V({\bf r})$
& $H_0=\frac{1}{2m}\left( -i\hbar\frac{\partial}{\partial {\bf r}}+
e{\bf A}_0({\bf r})\right)^2+V({\bf r})$\\
& & \\
Translation operators
&$T({\bf R})=e^{{\bf R}\cdot \partial/\partial {\bf r}}$
& ${\tilde T}({\bf R})=e^{ie/\hbar\int^{\bf R}_{\bf 0}d{\bf r}'\cdot {\bf
A}_0({\bf r}+{\bf r}')}
e^{{\bf R}\cdot \partial/\partial {\bf r}}$ \\
& & \\
Number of plaquettes             &1 plaquette                      &$q$
plaquettes \\
per unit cell& &\\
& & \\
Range of ${\bf k}$ vector & One Brillouin zone    & One magnetic Brillouin zone
\\
& & (One Brillouin zone divided by $q$) \\
& &\\
Perturbing fields  & ${\bf E,B}$  & ${\bf E},\delta {\bf B}$  \\
& &\\
Velocity of electron             & ${\bf\dot r}=\partial{\cal E}_n({\bf
k})/\hbar\partial {\bf k}$
&${\bf{\dot r}}=\partial E_n({\bf k})/\hbar\partial
{\bf k}-{\bf\dot k}\times {\bf \Omega}_n({\bf k})$, \\
& & $E_n({\bf k})={\cal E}_n^{\rm mag}({\bf k})+(e/2m)\delta {\bf B}\cdot {\bf
L}_n({\bf k})$\\
& &\\
Dynamics for ${\bf k}$
& $\hbar{\bf \dot k}=-e{\bf E}-e{\bf \dot r}\times {\bf B}$
& $\hbar{\bf \dot k}=-e{\bf E}-e{\bf \dot r}\times \delta {\bf B}$ \\
& &\\
Quantization condition & ${\rm
Area}(C_m)=2\pi\left(m+\frac{1}{2}\right)eB/\hbar$
& ${\rm Area}(C_m)=2\pi\left(m+\frac{1}{2}
-\frac{\Gamma(C_m)}{2\pi}\right)e\delta B/\hbar$ \\
for cyclotron orbits& &
\end{tabular}
\end{table}

\begin{table}
\caption{Hofstadter spectrum (for $\phi=22/67$) and the cyclotron energies
calculated from
Eq.~(4.5). Only the top ten subbands for the middle parent band are shown. The
last column is
the subband closest to the parent band edge ${\cal E}_2({\bf 0})=0.7321$.}
\begin{tabular}{ccccccccccc}
$E_{\rm Hofst}$ & 0.0618 & 0.1067 & 0.1566 & 0.2124 & 0.2747 & 0.3443 &
0.4221 & 0.5098 & 0.6098 & 0.7266 \\
& 0.0678 & 0.1085 & 0.1570 & 0.2125 & & & & & &  \\
$E_{\rm cyclo}$ & 0.0632 & 0.1063 & 0.1558 & 0.2115 & 0.2738 & 0.3435 & 0.4212
&
0.5086 & 0.6082 & 0.7240 \\
\end{tabular}
\end{table}

\begin{references}
\bibitem{Changmm}
In general, the magnetic field can be periodic.
This would only change the band structure, but not the dynamics itself.
See M.~C.~Chang and Q.~Niu, {\it Phys.~Rev.} {\bf B50}, 10843 (1994).
\bibitem{gauge}
Notice that since ${\bf r}_c$ is slowly varying with time (in the adiabatic
sense),
the gauge condition is also slowly changing with time.
\bibitem{odds}
Similar analysis of the oscillating part in ${\dot y}$ shows
that the correction from $\Omega$ does not lead to a net drift in the
$y$ direction.
\bibitem{even}
Another consequence of Eq.~(5.5) is that $q_r$ and $q_{r+1}$ cannot
both be even, for otherwise there will be a common factor of 2 for both
numerator and denominator of $p_{r+1}/q_{r+1}-p_{r}/q_{r}$.
This implies that the total number of daughter bands cannot be even if the
total number
of parent bands is even.
\bibitem{edge}
It is not $(4,-3,-3,4,-3,-3,4)$ because the subbands near the edges always have
$\sigma_{\rm close}$.
\end{references}
\end{document}